\documentclass[doublecol,figure]{epl2} 
\title{A Monte Carlo Method for Modeling Thermal Damping: Beyond the Brownian-Motion Master Equation} 

\author{Kurt Jacobs} 
\shortauthor{K. Jacobs} 

\institute{Department of Physics, University of Massachusetts at Boston, 100 Morrissey Blvd, Boston, MA 02125, USA}

\abstract{The ``standard" Brownian motion master equation, used to describe thermal damping, is not completely positive, and does not admit a Monte Carlo method, important in numerical simulations. To eliminate both these problems one must add a term that generates additional position diffusion. He we show that one can obtain a completely positive simple quantum Brownian motion, efficiently solvable, without any extra diffusion. This is achieved by using a stochastic Schr\"{o}dinger equation (SSE), closely analogous to Langevin's equation, that has no equivalent Markovian master equation.  Considering a specific example, we show that this SSE is sensitive to nonlinearities in situations in which the master equation is not, and may therefore be a better model of damping for nonlinear systems.} 

\pacs{05.40.Jc}{Brownian motion}
\pacs{03.65.Yz}{Decoherence; open systems; quantum statistical methods}
\pacs{03.65.Ta}{Foundations of quantum mechanics; measurement theory}
\pacs{85.85.+j}{Micro- and nano-electromechanical systems (MEMS/NEMS) and devices}

\begin{document}

\newtheorem{theo}{Theorem}
\newtheorem{lemma}{Lemma} 
%@twocolumnfalse\endcsname

\maketitle

All mechanical oscillators experience frictional damping, in which they lose energy to their  environment. This damping is accompanied by thermalization of the oscillator, since the environment is a large system, and thus a thermal bath. One would naturally like to model the effects of this damping on the oscillator, and indirectly on any other devices to which it is coupled, without having to describe the motion of the environment, with its many degrees of freedom. Frictional damping of quantum systems by an environment, often referred to as ``quantum Brownian motion", has many applications. In particular, it is essential in describing the behavior of nano-mechanical resonators, and is therefore of current interest in quantum nano-electro-mechanics (QNEMS)~\cite{Blencowe04, Naik06} and quantum opto-mechanics~\cite{Gigan06, Thompson08}. For classical systems, Langevin's equation~\cite{Langevin08}, containing only a deterministic frictional force and a Gaussian white noise source, gives a simple and excellent model of damping and thermalization. The situation is much less simple for quantum systems, however. 

The standard approach to obtaining a quantum equation to model damping from a thermal bath is to derive a master equation for a linear oscillator by coupling it to a  continuum of oscillators, and then trace out these oscillators. This was first done by Caldeira and Leggett~\cite{Caldeira83}, and was brought to its completion in the tour-de-force by Hu {\em et al.}~\cite{Hu92}, building on the work of Unruh and Zurek~\cite{Unruh89}. They showed that an exact master equation for the oscillator alone could be derived, and that any non-markovian effects of the bath appear merely as time-dependent coefficients in this equation. This master equation, the exact BME, is~\cite{Hu92} 
\begin{equation}
   \dot{\rho} = - i \Gamma(t) [x, \{ p , \rho \}]  - \xi(t) [x,[x,\rho]]  + \zeta(t) [x,[p,\rho]] .  
   \label{genBME}
\end{equation}
Here $\rho$ is the density matrix for the oscillator, $\{ p , \rho \} = p\rho + \rho p$ is the anti-commutator, and $x$ and $p$ are the dimensionless position and momentum of the oscillator, defined as 
\begin{eqnarray}
   x & = & (a + a^\dagger)/\sqrt{2} , \\ 
   p & = & -i(a - a^\dagger)/\sqrt{2} ,  
\end{eqnarray}
where $a$ is the oscillator's annihilation operator. The three functions of time, $\Gamma$, $\xi$ and $\zeta$, are the time-dependent coefficients that encode all non-Markovian effects of the bath, and are determined by the frequency of the resonator, $\omega$, the coupling to the bath (which sets the damping rate, $\gamma$), the structure of the bath (ohmic, super-ohmic, etc.), and the initial bath state (which is determined by the temperature, $T$). For reference, the physical position and momentum are $X = \sqrt{\hbar/(\omega m)} x$ and $P = \sqrt{\hbar \omega m} p$, where $m$ is the mass and $\omega$ the frequency of the oscillator.  We note that the master equation that results from using a bath of two-level systems, rather than oscillators, has the same form as Eq.(\ref{genBME})~\cite{TLbath}. 

There are three issues with using the BME to model damping. The first is that the time-dependent coefficients are in general rather complex, and must be calculated numerically for a given problem~\cite{Hu92}. The second is that this equation cannot be written in the Lindblad form~\cite{Lindblad76}. Because of this it has no equivalent stochastic Schr\"{o}dinger equation (SSE, also referred to as an ``unravelling")~\cite{JacobsSteck06}, and therefore cannot be simulated using the wave-function Monte Carlo method~\cite{Molmer93}. This method significantly reduces the numerical overhead involved in simulating master equations, and is therefore of considerable practical  importance. Thirdly, this master equation only generates the correct thermal steady-state for linear oscillators, and many questions of interest for mesoscopic oscillators involve non-lineairites of the oscillators themselves, or interactions with other nonlinear devices (such interactions effectively turn an oscillator into a nonlinear system). As far as the author is aware, no master equations have been derived for damping of nonlinear oscillators. Instead, the problem is avoided: since it is usually the case in practice that the nonlinearities are small compared to the harmonic motion, one assumes that the BME (and the SSE to be presented shortly) will give reasonable results. Ideally one would like to improve upon this situation, however, and with this in mind we consider the behavior of various BME variants, as well that of the SSE, for an example nonlinear system in detail below. 

The standard approach to addressing the first issue is to choose constant values for the time-dependent coefficients, so as to keep the resulting BME accurate to a good approximation. Different choices for these values produce different variants. For oscillators at low temperatures, the relevant regime for QNEMS, a reasonable approximate version is~\cite{Caldeira89}  
\begin{equation}
   \dot{\rho} =  - i \frac{\gamma}{4} [x, \{ p , \rho \}]  - \frac{\gamma}{4} ( 2 n_T + 1 ) [x,[x,\rho]]  .   
   \label{lbme}
\end{equation}
Here $n_T$ is the average number of phonons in the resonator when it is at temperature $T$, being  
\begin{equation}
   n_T = \frac{1}{ \exp\left[\hbar \omega/(k_{\mbox{\scriptsize B}} T)\right] - 1} , 
   \label{nt} 
\end{equation}
with $k_{\mbox{\scriptsize B}}$ Boltzmann's constant. Note that in what follows, $n_T$ will always be defined as the specific function of $T$ given in Eq.(\ref{nt}). We will refer to Eq.(\ref{lbme}) as the ``low temperature BME" (LBME). For completeness we note that a similar variant, sometimes refereed to in the quantum optics literature as the ``standard" BME (SBME), is given by 
\begin{equation}
   \dot{\rho} = - i \frac{\gamma}{4} [x, \{ p , \rho \}]  
    - \gamma \left( \frac{k_{\mbox{\scriptsize B}} T}{2 \hbar \omega} \right) [x,[x,\rho]]  .    
\end{equation}  
This SBME is a reasonable approximation to the BME in the joint regime of high temperature {\em and} weak damping~\cite{Caldeira89}. These two variants still have no equivalent SSE. They also generate a non-physical term in the fluctuation spectrum of the resonator~\cite{Jacobs99} (due to the fact that the evolution is not completely positive~\cite{Munro96,Gnutzmann96}), although this does not usually cause problems in the time domain. 

As first pointed out by Diosi~\cite{Diosi93}, one can obtain a new BME that {\em does} have an equivalent SSE by adding a term to the LBME or SBME. The low-temperature version, which we will refer to here as the ``completely positive" BME (PBME) is 
\begin{eqnarray}
   \dot{\rho} & = &  - i \frac{\gamma}{4} [x, \{ p , \rho \}] - \frac{\gamma}{4} ( 2 n_T + 1 ) [x,[x,\rho]]  \nonumber \\ 
                   & &    - \frac{\gamma}{ 16 ( 2 n_T + 1 )} [p,[p,\rho]] .  
                   \label{pbme}
\end{eqnarray} 
The extra term generates diffusion in position. This equation is a considerable improvement, as it is now a valid quantum evolution, and can be simulated using the wave-function Monte Carlo method. However, the PBME no-longer gives the correct thermal steady-state for a linear oscillator. This is because the steady-state average phonon number is increased by the position diffusion. Note that the position diffusion term in the PBME is {\em not} part of the exact BME derived by Hu {\em et al.} Nevertheless, a master equation of the form given in Eq.(\ref{pbme}) can be derived as an approximation to the dynamics induced by a collisional bath (rather than a linear bath of harmonic oscillators)~\cite{Diosi95, Barnett05, Oppenheim87, Petruccione05, Hornberger06, Hornberger07}), and the high temperature version of Eq.(\ref{pbme}) has also been derived as an approximate limit of a system interacting with a bath of oscillators~\cite{Presilla97}. 

We now present a quite different approach to modeling damping and thermal noise in mesoscopic oscillators. This involves directly constructing a stochastic Schr\"{o}dinger equation for this purpose, rather than first deriving a master equation. In fact, while the resulting SSE is straightforward to simulate~\cite{code}, it does not correspond to {\em any} Markovian master equation. The SSE satisfies all the properties required for the thermal damping of a linear system, in particular generating the correct thermal steady-state. Since this equation has no additional position diffusion, it can be regarded as a genuine quantum version of simple damped Brownian motion. Further the motion of the mean values of position and momentum for a given trajectory are precisely those given by Langevin's equation (They contain a fluctuating force that generates momentum diffusion.) The SSE may therefore be regarded as a quantum equivalent of Langevin's equation. Performing simulations of a particular example, we show that while the SSE does not produce the correct (thermal) steady-state for nonlinear systems, it does better than the various versions of the Brownian motion master equation.  

The SSE we propose to model thermal damping is  
\begin{equation}
    \mbox{see Eq.(9)} \nonumber. 
\end{equation}
\begin{widetext} 
\begin{eqnarray}
  d |\psi\rangle & = &  \left[ - i \frac{H}{\hbar} - \, \frac{\gamma  n_T}{2} \left( x^2 - p^2 \right) \right] |\psi\rangle \; dt \, + \, 2 \gamma \left(  n_T \langle x \rangle x +  n_T \langle p \rangle p  + \frac{i}{4}  \langle p \rangle x  \right) |\psi\rangle \; dt  \nonumber \\
                       & & + \, 2\sqrt{\gamma n_T} \left( \frac{x}{2}  + iV_x p - i \left[ C_{xp} - \frac{1}{2}\right] x \right)   |\psi\rangle \;  dW_1 \, + \, 2 \sqrt{\gamma n_T} \left( \frac{p}{2}  - iV_p x + iC_{xp} p \right)  |\psi\rangle \;  dW_2 . 
                       \label{sbmsse} 
\end{eqnarray}
\end{widetext} 
Here $H$ is the oscillator Hamiltonian, and the increments $dW_1$ and $dW_2$ are mutually independent Wiener noise increments satisfying the Ito calculus relation $dW_1^2 = dW_2^2 =  dt$~\cite{WienerIntroPaper,JacobsSteck06}. As is usual, the means, variances, and covariance of the position and momentum for the state $|\psi\rangle$ are denoted by $\langle x \rangle = \langle \psi | x | \psi \rangle$, $\langle p \rangle = \langle \psi | p | \psi \rangle$, $V_x = \langle x^2\rangle - \langle x \rangle^2$, $V_p = \langle p^2\rangle - \langle p \rangle^2$, and $C_{xp} = \langle xp + px\rangle/2 - \langle x \rangle \langle p \rangle$. 

To simplify the presentation, the above SSE is {\em unnormalized}~\cite{JacobsSteck06}. This means that to simulate the SSE, at each time-step $dt$ one adds the increment $d|\psi\rangle$, and then must {\em normalize} the state before adding the next increment. To obtain the density matrix for an oscillator undergoing Brownian motion, one simulates the above equation for many realizations of the noise processes $dW_i$. If we label the states resulting from each of the $N$ noise process as $|\psi_j(t)\rangle$, $j=1,\ldots,N$,  then the density matrix for the system at time $t$ is $\rho(t) = (1/N)\sum_{j} |\psi_j (t) \rangle \langle \psi_j (t) |$.   

The SSE (Eq.~(\ref{sbmsse})) has the following properties: 

1. It gives a valid quantum evolution. 

2. It gives the Langevin equations for the mean values of position and momentum: 
\begin{eqnarray}
    d\langle x \rangle & = &  -i \langle [  x , H ] \rangle dt ,  \label{eqm1} \\
    d\langle p \rangle & = &  -i  \langle [ p, H ]  \rangle dt  - \frac{\gamma}{2} \langle p \rangle dt  + \sqrt{\gamma  n_T} dW_2 .   \label{eqm2}
\end{eqnarray} 

3. For a harmonic oscillator, the SSE reduces every initial state to a coherent state (with randomly fluctuating mean position and momentum) as $t\rightarrow\infty$. 

The first two properties confirm that the SSE realizes a quantum equivalent of classical damped Brownian motion. The third shows that this evolution also gives the same final states as the standard Brownian motion master equation. Together these show us that the SSE provides a good description of simple damped quantum Brownian motion. The final property is now expected to follow, and we show below that it does:

4. For a harmonic oscillator the SSE gives the expected thermal steady-state density matrix,  being  
\begin{equation}
   \rho_{\mbox{\scriptsize ss}} = \left( \frac{1}{ n_T+1} \right)\sum_{n=0}^{\infty} \left( \frac{ n_T}{ n_T+1} \right)^n |n\rangle \langle n | , \label{rhoss}
\end{equation}
where the states $|n\rangle$ are the phonon number states of the resonator. This can also be written as a Gaussian Wigner function. The steady-state covariances are  
\begin{equation}
   V_x^{\mbox{\scriptsize ss}} =  V_p^{\mbox{\scriptsize ss}} = \frac{1}{2} +  n_T ,
\end{equation}
and $C_{xp}^{\mbox{\scriptsize ss}} = 0$. 

{\em Constructing the SSE:} 
To obtain an SSE with the four properties listed above, we begin by noting that for a linear oscillator, if we make two simultaneous continuous measurements, one of $x$ and the other of $p$, this gives an SSE whose steady-state is a coherent state with randomly fluctuating mean position and momentum. The SSE, normalized this time, for this joint measurement is~\cite{JacobsSteck06} 
\begin{equation}
    \mbox{see Eq.(15)} \nonumber. 
\end{equation}
\begin{widetext} 
\begin{eqnarray}
   d |\psi\rangle & = &  - k \left[ (x - \langle x \rangle)^2 + (p - \langle p \rangle)^2 \right] dt |\psi\rangle + \sqrt{2k} \left[ (x - \langle x \rangle) dW_1 + (p - \langle p \rangle) dW_2  \right] |\psi\rangle 
      \label{jointmeas}
\end{eqnarray}
\end{widetext} 
where $k$ determines the rate at which the measurement extracts information (often called the ``measurement strength").  The equations of motion for the means that result from this SSE are 
 \begin{eqnarray}
   d \langle x \rangle & = & \sqrt{8k}V_x dW_1 +  \sqrt{8k}C_{xp} dW_2 , \\ 
   d \langle p \rangle & = & \sqrt{8k}V_p dW_2 +  \sqrt{8k}C_{xp} dW_1 , 
\end{eqnarray}
and the equations for the covariances, when the wavefunction for the state $|\psi\rangle$ is Gaussian, are 
\begin{eqnarray}
  \dot{V}_x   & = & -8k V_x^2 - 8k C_{xp}^2 + 2 k   , \label{eqv1} \\ 
  \dot{V}_p   & = & -8k V_p^2 - 8k C_{xp}^2 + 2 k  , \label{eqv2} \\ 
  \dot{C}_{xp} & = & -8k (V_x + V_p) C_{xp}  .   \label{eqv3}
\end{eqnarray}
For Brownian motion we require that the mean position is not driven by any noise, and that the mean momentum has the equation $d\langle p \rangle = -(\gamma/2) \langle p \rangle dt + \sqrt{\gamma n_T} dW_2$. We can produce the correct equations for the means by adding a linear feedback term, applied by a fictitious observer who has access to the stream of measurement results from the continuous measurements of $x$ and $p$~\cite{DJ}. We choose the feedback Hamiltonian to cancel the noise driving the mean position, and to add damping to the mean momentum. The required feedback Hamiltonian is 
\begin{eqnarray}
  H_{\mbox{\scriptsize fb}} \! & = & \! 2 \hbar \sqrt{\gamma n_T } \left(  V_p \frac{dW_2}{dt} + \left[C_{xp} - \frac{1}{2}\right]  \frac{dW_1}{dt}  \right) x  \;\;\;\;\;\;\;\; \nonumber \\ 
                              & & + \frac{\gamma \langle p \rangle}{2} x - 2 \hbar  \sqrt{\gamma n_T } \left(  V_x \frac{dW_1}{dt}  +C_{xp} \frac{dW_2}{dt}   \right) p   \nonumber .
\end{eqnarray}
With the addition of this fictitious feedback, there is only one more thing we need to do to obtain the Brownian motion SSE. This is to fix the value of the measurement strength, $k$. Since it is the thermal bath that causes both the measurement (decoherence) and the damping, $k$ should be related to the damping rate, $\gamma$. To get this relationship right, we thus choose $k$ so that the quadratic terms in our SSE (the terms that give the decoherence due to the measurement) match those of the simple forms of the standard Brownian motion master equation that give the same damping rate~\cite{Diosi95,Caldeira89}. (These quadratic terms also match the Lindblad master equation for the damping of an optical cavity with the same damping rate~\cite{QuantumNoise}). This sets $k = (\gamma n_T)/2$, and results in the SSE given by Eq.(\ref{sbmsse}). 

By construction, the SSE gives a valid (completely positive) quantum evolution (property 1 above), and the equations of motion for the expectation values of $x$ and $p$ satisfy  property 2. It is well-established that continuous measurement of any linear combination of $x$ and $p$ produces Gaussian states as $t\rightarrow\infty$~\cite{Doherty99, Bhattacharya00}. While to the authors knowledge no formal proof of this has been given to date, the proof is fairly straightforward, and follows from the results in~\cite{JSinprep}. Once the wave-function is Gaussian, the equations of motion for the covariances are those given in Eqs.(\ref{eqv1})-(\ref{eqv3}), with the addition of the harmonic oscillator dynamics, given by 
\begin{eqnarray}
  \dot{V}_x  & = & 2 \omega C_{xp} , \label{heqv1} \\ 
  \dot{V_p}  & = & - 2 \omega C_{xp} , \label{heqv2} \\ 
  \dot{C}_{xp} & = & \omega (V_p - V_x)  .   \label{heqv3}
\end{eqnarray}
The the linear feedback Hamiltonian has no effect on the covariances. The resulting steady-state is $V_x = V_p = 1/2$, so the SSE satisfies property 4. 

To show the the SSE satisfies property 3, we first note that for a harmonic oscillator, in the long-time limit, the density matrix is a mixture of coherent states, due to property 4. The steady-state density matrix can therefore be obtained by calculating the probability density for $\langle x \rangle$ and $\langle p \rangle$ generated by Eqs.(\ref{eqm1}) and (\ref{eqm2}). Solving these equations is straightforward, and the resulting steady-state probability density for $\langle x \rangle$ and $\langle p \rangle$ is a Gaussian with zero mean and variances $V_{\langle x \rangle} = V_{\langle p \rangle} = n_T$. The steady-state for the system is therefore a Gaussian mixture of coherent states, and has the variances $V_x = V_p = 1/2 + n_T$. This is precisely the thermal state given by Eq.(\ref{rhoss}). 

As mentioned above, the SSE does not correspond to any Markovian master equation. The reason for this is the feedback. Without feedback one can average the SSE over many trajectories and obtain a Markovian master equation for the density matrix of the system. However, with feedback the dynamics of each trajectory depends on the state of the system during that trajectory in a way that cannot be reduced to a function of the density matrix (being the average over all trajectories) at the current time. The dynamics  of the density matrix at time $t$ cannot therefore be written purely in terms of the density matrix at $t$, making a description in terms of a Markovian master equation impossible. 

\begin{figure}[t] 
\leavevmode\includegraphics[width=1\hsize]{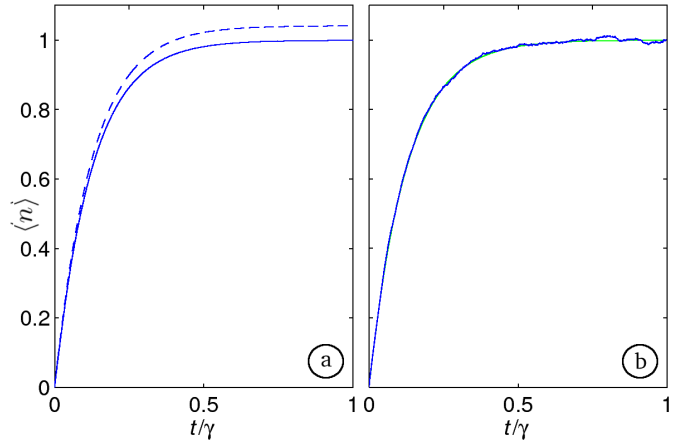} 
\caption{Here we show the evolution of the mean phonon number of a harmonic oscillator, $\langle n \rangle$, under three evolutions that model damping and thermalization (Brownian motion). The temperature parameter in the equations is chosen so that the correct thermal steady-state value of $\langle n \rangle$ is 1. a) Solid line: the low temperature Brownian motion master equation (LBME); Dashed line: the completely positive version of the same master equation. b) Dark solid line: The stochastic Schr\"{o}dinger equation described here, averaged over approximately 20,000 trajectories; Light Solid Line: The result for the LBME, which is largely obscured behind.} 
\label{fig1} 
\end{figure}

{\em Damping of Nonlinear systems:} We will now consider the behavior of the SSE for a particular example of a nonlinear oscillator, and compare this with the behavior of the BME variants. As a reference we first solve the LBME and PBME master equations numerically, along with the SSE, for a linear oscillator. Measuring time in units of the oscillation period of the oscillator, we have $\omega = 2 \pi$. We further choose the damping rate $\gamma = 4$, the temperature so that $n_T = 1$ (this means that $T = k_{\mbox{\scriptsize B}}/(\hbar \omega \ln 2)$), and start the oscillator in the ground state ($n=0$).  In Fig.~\ref{fig1} we plot the time evolution of the mean phonon number under the two master equations and the SSE. The evolution of the SSE is indistinguishable from that of the LBME, both of which settle down to the expected steady-state value of the average phonon number, $\langle n \rangle = n_T = 1$.  

We now consider a $\chi^{(3)}$ nonlinear oscillator, which has the Hamiltonian~\cite{PRLinsub}
\begin{equation}
   H_{\mbox{\scriptsize nl}}  = \hbar \omega (a^\dagger a)^2 . 
\end{equation}
This Hamiltonian has the same eigenstates as the harmonic oscillator, $|n\rangle$, $n = 0, 1, 2,\ldots$, but the energy levels are now $E_n = \hbar\omega n^2$. It is only the two lowest energy levels that the $\chi^{(3)}$ oscillator has in common with the harmonic oscillator (up to a shift of $\hbar\omega/2$).  

Simulating the nonlinear oscillator with the two master equations, with $\omega$, $\gamma$ and $T$ (and thus $n_T$) as before, one learns immediately that these give  the same steady-state distribution over the eigenstates, $|n\rangle$, as they do for the linear oscillator, and thus the same average value of $n$. That is, so long as the eigenstates are those of the harmonic oscillator, the steady-states of these master equations know nothing about the energy levels of the system. This behavior is easily understood: the master equations eliminate the off-diagonal elements of the density matrix in the energy eigenbasis, and once this has happened the Hamiltonian has no effect on the evolution. Since the energy eigenvalues only act on the dynamics via the Hamiltonian, they can have no effect on the steady-state.  

The average energy for the nonlinear oscillator is no longer given by $\hbar\omega (\langle n \rangle + 1/2)$. The energy levels are now proportional to $n^2$, and the average energy is $\langle E \rangle = \hbar\omega \langle n^2 \rangle$. The steady-state distribution for the nonlinear oscillator given by the LBME is not the thermal (Boltzmann) distribution. As noted above it remains the thermal distribution for the harmonic oscillator, and therefore gives too much weight to the higher energy levels. For $T =  k_{\mbox{\scriptsize B}}/(\hbar \omega \ln 2)$ (that is, $n_T = 1$), the thermal value of $\langle n^2 \rangle$ for the nonlinear oscillator is $0.49$, and that given by the LBME is $3$. To use the LBME to model damping of the nonlinear oscillator at a given temperature, we could always alter the parameter $n_T$ so that the LBME gives the correct thermal value for $\langle n^2 \rangle$ at that temperature (and thus for the average energy). Of course this does not achieve the Boltzmann distribution.  

\begin{figure}[t] 
\leavevmode\includegraphics[width=1\hsize]{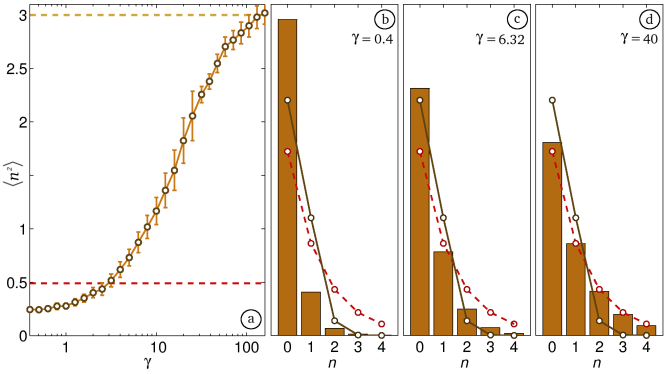} 
\caption{Here we show the steady-state mean energy, $\langle E \rangle \propto \langle n^2 \rangle$, for a $\chi^{(3)}$ nonlinear oscillator, as well as the steady-state probability distribution for the energy levels, $n$, for two evolutions: the low temperature Brownian motion master equation, and the SSE described here. The momentum damping rate is $\gamma/2$, and the temperature is chosen so that $n_T=1$, and a harmonic oscillator with the same lowest two energy levels has $\langle n^2 \rangle=3$. (a) circles: the value of $\langle n^2 \rangle$ given by the SSE; dark dashed line: thermal value for $\langle n^2 \rangle$; light dashed line: $\langle n^2 \rangle = 3$. (b)-(d): The distribution over the energy levels for three damping rates. Bars: the distribution given by the SSE; solid line: the thermal (Boltzmann) distribution; dashed line: the distribution given by the Brownian motion master equations.} 
\label{fig2} 
\end{figure} 

The behavior of the SSE for the nonlinear oscillator is quite different to that of the master equations. The SSE is sensitive to the energy levels of the oscillator, and the weighting of higher energy levels is suppressed as one would like. However, it also does not achieve the Boltzmann distribution. In addition, the steady-state value of $\langle n^2\rangle$, while usually closer to the thermal value than that given by the master equations, depends on the damping rate. In Fig.~\ref{fig2}(a) we plot the steady-state value of $\langle n^2 \rangle$ as a function of the damping rate, with the parameter $n_T = 1$ as before. For large damping the distribution given by the SSE tends to that given by the master equation (being the Boltzmann distribution for the harmonic oscillator). As the damping rate is reduced, $\langle n^2 \rangle$ decreases, and, after crossing the thermal value, settles into a limiting value, becoming independent of the damping rate for weak damping. For $n_T = 1$, the SSE gives the thermal value for $\langle n^2 \rangle$ when $\gamma \simeq 0.8$. For weak damping the SSE gives $\langle n^2 \rangle = 0.24$ which is almost exactly half the thermal value. 

In Fig.~\ref{fig2} (b)-(d) we show the steady-state distribution over the energy levels given by the SSE for three values of the damping, and compare this to the correct thermal distribution for the nonlinear oscillator, and the thermal distribution for the harmonic oscillator. We see that for small damping, the SSE actually does the opposite of the LBME, in that it weights the lower energy levels {\em more} than the thermal distribution, and thus underestimates the average energy. As the damping increases, the distribution changes and tends towards the thermal distribution for the harmonic oscillator, thus giving the same result as the LBME in the limit of large damping. We see from Fig.~\ref{fig2} (c) that at no point does the distribution match the thermal distribution: when the average energy is equal to the thermal value, the SSE weights both $n<1$ and $n>1$ more than the thermal distribution.  

{\em The case of distant systems:} It is worth noting that there is a wrinkle to using the SSE for simulating damping on systems that are entangle with distant systems, and in which one wishes explicitly to examine questions of non-locality. In order to provide the feedback required to generate the damping, the bath (the fictitious observer) must have maximal knowledge of the state of the system. This is already implied by the fact that the SSE propagates a pure state, and causes no issue if the system is coupled only to other local systems. However, if one wishes to describe a thermal resonator that is entangled with a second, distant system, then the fact that we propagate a pure state for the resonator and the second system means that we give the bath full knowledge of this second system. This is also fine, unless we have an observer performing operations on the distant system while the resonator is evolving. In this case, allowing the bath access to the joint pure-state of both systems means that we are assuming infinitely fast communication from the distant observer to the local bath. 

If one does wish to use the SSE in such a situation, then one can preserve locality by modifying the simulation in the following way. One first includes the measurements of the distant observer in the SSE. Secondly, one averages over these measurements to obtain a stochastic master equation (SME) that gives the state of knowledge of the bath. The feedback applied by the bath is now determined by the state-of-knowlegde of the bath, and not by the SSE. The SSE and SME are then propagated together, both containing the feedback terms that are now determined by the SME. This procedure preserves locality since the bath, and thus the feedback, only ever has access to local information. However, it should be noted that doing this {\em does} modify the feedback, and therefore the motion induced by the bath. That is, the motion is no-longer exactly the simple Brownian motion that we wish to simulate, at least while the state-of-knowledge of the bath remains appreciably mixed. The SSE is therefore not always appropriate for  problems involving explicit investigations of nonlocality. 

{\em Conclusion:} We have shown how to obtain a quantum model of simple damped Brownian motion that gives a completely positive evolution. This is achieved by constructing a stochastic Schr\"{o}dinger equation rather than a master equation, and by using the physical notion of feedback from a measurement process, a concept borrowed from the subject of quantum feedback control~\cite{DJ}. 

While both the SSE and the Brownian motion master equations give the correct thermal distribution for the harmonic oscillator, interesting nano-mechnical devices are usually nonlinear, even if relatively weakly. We therefore explore using the LBME and the SSE to model damping of an oscillator in an extreme case, in which the Hamiltonian is a $\chi^{(3)}$ nonlinearity. While the LBME is completely insensitive to the energy levels of this system (and thus, completely insensitive to the nonlinearity) the SSE is not. For weak damping the distribution given by the SSE is sensitive to the nonlinearity (and independent of the damping rate) although it {\em overestimates} the probabilities of the lower energy levels. In the limit of strong damping it reproduces the results of the LBME. For intermediate damping the distribution of the energy levels, and thus the average energy, is dependent upon the damping rate. These results suggest that the SSE is certainly no worse, and probably a better, model of thermal damping for nonlinear systems than the various forms of the Brownian motion master equation. 

The present work highlights the fact that current techniques for modeling damping and thermalization in nonlinear systems are poor. It may be that stochastic Schr\"{o}dinger equations, and the concept of feedback, will be a useful tool in developing more accurate models that are also efficient to implement. 

{\em Acknowledgments:} I thank Lajos Diosi for stimulating discussions, and for raising the question of simulating Brownian motion with distant systems. I am also indebted to Salman Habib for many enlightening discussions regarding Brownian motion over a number of years. This work was performed with the supercomputing facilities in the School of Science and Mathematics at UMass Boston. 

%\bibliographystyle{apsrev} 
%\bibliography{report}

\begin{thebibliography}{27}
\bibitem{code} 
  Code for simulating the SSE using a spectral split-operator method 
  can be obtained free of charge from the author's website.
\end{thebibliography}

\end{document}